\documentclass[aps,prl,showpacs,twocolumn,a4paper,10pt]{revtex4}

\usepackage{amssymb}
\usepackage{amsmath}
\usepackage{graphicx}

\begin{document}

\title{Dynamical Creation of Fractionalized Vortices and
Vortex Lattices}
\author{An-Chun Ji$^1$, W. M. Liu$^1$,
Jun Liang Song$^2$ and Fei Zhou$^2$} \affiliation{$^1$Beijing
National Laboratory for Condensed Matter Physics, Institute of
Physics, Chinese Academy of Sciences, Beijing 100080,
China \\
$^2$Department of Physics and Astronomy, The University of British
Columbia, Vancouver, B. C., Canada V6T1Z1}
\date{{\small \today}}

\begin{abstract}
We investigate dynamic creation of fractionalized half-quantum
vortices in Bose-Einstein condensates of sodium atoms. Our
simulations show that both individual half-quantum vortices and
vortex lattices can be created in rotating optical traps when
additional pulsed magnetic trapping potentials are applied. We also
find that a distinct periodically modulated spin-density-wave
spatial structure is always embedded in {\em square} half-quantum
vortex lattices. This structure can be conveniently probed by taking
absorption images of ballistically expanding cold atoms in a
Stern-Gerlach field.
\end{abstract}


\maketitle

Topological excitations such as quantized vortices have been
fascinating for quite a few decades and recently have also been
thoroughly studied in Bose-Einstein condensates (BECs) of ultracold
atoms \cite{Matthews99,Madison00,AboShaeer01,Haljan01,Fetter01,
Dalfovo01,Tsubota02,Lundh03,Parker05}. For vortices in single
component BECs, the circulation of supercurrent velocity ${\bf v}_s$
along a closed curve $\Gamma$ around a vortex line (defined as
${\cal C}=\int_{\Gamma} d{\bf l} \cdot {\bf v}_s$) is quantized in
units of $2\pi \hbar/m$ ($m$ is the atomic mass), with ${\cal C}=\pm
1, \pm 2, ...$ as a consequence of {\em analyticity} of
single-valued wavefunctions of coherent quantum states. Furthermore,
only a vortex with circulation ${\cal C}=\pm 1$ or an elementary
vortex is energetically stable. A secondary vortex with circulation
${\cal C}=\pm 2, \pm 3,...$ spontaneously splits into a few
elementary ones which interact via long range repulsive potentials.

A configuration with its circulation smaller than the elementary
value (${\cal C}=1$) has to be described by a singular wavefunction
and it always turns out to be energetically catastrophic. A most
obvious example is a two-dimensional configuration, where the
condensate phase angle $\Phi(r,\theta)$ rotates slowly and uniformly
by $180^\circ$ in the $r-\theta$ plane around a vortex center but
jumps from $\pi$ to $2\pi$ when the polar angle $\theta$ is equal to
$2\pi$. The $\pi$-phase jump here effectively induces a singular
{\em cut} in the wavefunction. The corresponding circulating
velocity field is simply ${\bf v}_s(r,\theta)=\hbar/(2 m r) {\bf
e}_\theta$, leading to ${\cal C}=1/2$ that is one-half of an
elementary value. The energy of a {\em cut} per unit length along
$\theta=2\pi$ line where the phase jumps is finite and therefore the
overall energy of a {\em cut} in an individual fractionalized vortex
scales as $L$, $L$ is the size of system, while the energy for an
integer vortex only scales as a logarithmic function of $L$.
Consequently, a {\em cut} that connects two singular fractionalized
vortices mediates a linear long rang {\em attractive} potential that
confines all fractionalized excitations. So in a single component
condensate, vortices with ${\cal C}=\pm 1$ are fundamental ones
which do not further split into smaller constituent elements as a
result of confinement of fractionalized vortices.

Hyperfine-spin degrees of freedom can drastically change the above
arguments about elementary vortices. In condensates of sodium
($^{23}Na$) or rubidium ($^{87}Rb$) atoms in optical traps
\cite{Stenger98}, hyperfine spins of cold atoms are correlated
because of condensation. A {\em pure} spin defect (or a spin
disclination in the case of $^{23}Na$) where spins of cold atoms
slowly rotate but no supercurrents flow, can carry a {\em cut}, i.e.
a line along which a $\pi$-phase jump occurs as a result of Berry's
phases induced by spin rotations \cite{Thouless98}. Such a spin
defect can then terminate a {\em cut} emitted from a singular
half-quantum vortex ({\em HQV}) configuration, which consequently
leads to a linear confining potential between the spin defect and
{\em HQV}. For instance, a {\em HQV} with ${\cal C}=1/2$ confined to
a spin defect does exist as a fundamental excitation in spin nematic
condensates \cite{Zhou01,Leonhardt00}. In this Letter we will
present the simulation of dynamical creation of fractionalized {\em
HQV}s in a rotating BEC of sodium atoms, and formulate an
experimental procedure for the realization of such exotic
topological excitations.

{\em Magnetic properties and energetics of half vortices} Unlike
conventional integer vortices, {\em HQV}s have very rich magnetic
structures. We will demonstrate that far away from its core a {\em
HQV} have vanishing local spin densities but is accompanied by
slowly rotating spin quadrupole moments; within the core, spin
densities are nonzero. The Hamiltonian for interacting sodium atoms
is
\begin{eqnarray}
H \!\!=\!\!\! \int\!\!\! d{\bf r} \psi^\dagger_{\alpha}({\bf r})
\frac{-\hbar^2\nabla^2}{2m} \psi_{\alpha}({\bf r}) \!+\!\!
\frac{c_2}{2}\!\! \int\!\!\! d{\bf r} {\hat{\bf S}}^2({\bf r})\!
+\!\! \frac{c_0}{2}\!\! \int\!\!\! d{\bf r} {\hat{\rho}}^2({\bf
r}).\label{Hamiltonian}
\end{eqnarray}
Here $\psi^\dagger_{\alpha}(\psi_{\alpha})$, $\alpha=x,y,z$ are
creation (annihilation) operators for sodium atoms in hyperfine
states $|\alpha\rangle$; they are defined as linear superpositions
of creation operators for three spin-one states, $|1, m_F\rangle$,
$m_F=0,\pm 1$.
$\psi^\dagger_x=(\psi^\dagger_{1}-\psi^\dagger_{-1})/\sqrt{2}$,
$\psi^\dagger_y=(\psi^\dagger_{1}+\psi^\dagger_{-1})/i \sqrt{2}$ and
$\psi^\dagger_z=\psi^\dagger_0$. $c_{0,2}$ are interaction
parameters that depend on two-body $s$-wave scattering lengths
$a_{0,2}$ for total spin $0$,$2$: $c_0=4\pi\hbar^2(a_0+2a_2)/3m$ and
$c_2=4\pi\hbar^2(a_2-a_0)/3m$. For condensates of $^{23}$Na atoms,
$a_0\simeq 50\>a_B$ and $a_2\simeq 55\>a_B$ ($a_B$ is the Bohr
radius). $\hat{\bf
S}_{\alpha}=-i\epsilon_{\alpha\beta\gamma}\Psi^*_\beta \Psi_\gamma$
and $\hat{\rho}=\psi^\dagger_{\alpha}\psi_{\alpha}$ are local
spin-density and density operators.

Generally, a condensate wavefunction
$\Psi_\alpha(=\langle\psi^\dagger_\alpha\rangle)$ for spin-one atoms
is a complex vector. For sodium atoms, interactions favor states
with a zero spin density; this leads to a spin nematic ground state
which does not break the time reversal symmetry and
$\Psi=\exp(i\Phi) \sqrt{\rho} {\bf n}$; here ${\bf n}$ is a unit
vector with three components $n_\alpha$, and $\rho$ is the number
density of sodium atoms. This spin nematic state is invariant under
an inversion of ${\bf n}$ and a $\pi$-phase shift (i.e. ${\bf n}
\rightarrow -{\bf n}$ and $\Phi \rightarrow \Phi +\pi$). Although
local spin densities $\langle\hat{\bf S}({\bf r})\rangle$ in a
nematic state vanish, a nematic condensate carries a spin quadrupole
moment defined as $Q_{\alpha\beta}({\bf r})=\langle \hat{\bf
S}_\alpha \hat{\bf S}_\beta\rangle-\frac{1}{3}\delta_{\alpha\beta}
\langle \hat{\bf S}^2\rangle$, which can be calculated and is
specified by the nematic unit director ${\bf n}$ introduced above:
$Q_{\alpha\beta}=\rho ({\bf n}_\alpha {\bf
n}_\beta-\frac{1}{3}\delta_{\alpha\beta}$).

Around a {\em HQV} centered at the origin and oriented along the
$z$-direction, nematic directors lying in a perpendicular $r-\theta$
plane rotate by $180^\circ$ forming a spin-disclination; the
corresponding condensate wavefunction far way from the vortex core
is $\Psi(\theta,r \rightarrow \infty)= \exp(i \theta/2 )\sqrt{\rho}
{\bf n}(\theta)$, $n_x\!\!=\!\!\cos(\theta/2)$,
$n_y\!\!=\!\!\sin(\theta/2)$ and $n_z\!\!=\!\!0$. The $180^\circ$
rotation of nematic director ${\bf n}(\theta)$ around the vortex
illustrates that a $\pi$-spin disclination where spin quadrupole
moments $Q_{\alpha\beta}({\theta})$ slowly rotate is indeed confined
to this {\em HQV}. In the Zeeman basis of $|1,m_F\rangle$ states,
$m_F=0,\pm 1$, the above vortex state is equivalent to
$\psi_{1}=\exp(i\theta) f(r) \sqrt{\rho}/\sqrt{2}$, $\psi_{-1}=-g(r)
\sqrt{\rho}/\sqrt{2}$ and $\psi_0=0$; far away from the vortex core,
$f(r\rightarrow \infty )=g(r\rightarrow \infty)=1$. The core
structure can be studied by numerically solving the
multiple-component Gross-Pitaevskii (GP) equation
\begin{eqnarray}
[-\frac{\hbar^2}{2m}\nabla^2 + (c_0+c_2)\rho_{\pm 1} +
(c_0-c_2)\rho_{\mp 1}]\psi_{\pm1}=0, \label{corestructure}
\end{eqnarray}
here $\rho_{m_F}=|\psi_{m_F}|^2$. The corresponding boundary
conditions when $r\rightarrow\infty$ are set by the asymptotic
behaviors of a {\em HQV} far away from the core as discussed before
Eq.(\ref{corestructure}). Notice that the last term in
Eq.(\ref{corestructure}) indicates that mutual interactions between
$|1,\pm 1\rangle$ atoms induced by scattering are repulsive since
$c_0 -c_2$ is positive.

\begin{figure}
\includegraphics[width=\columnwidth]{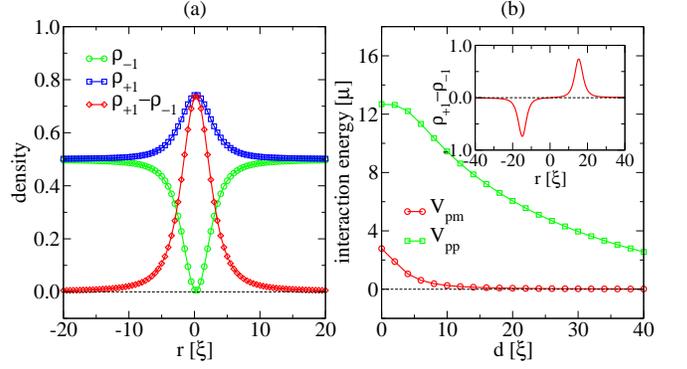}
\caption{(color online). (a) Density ($\rho_{\pm 1}=|\psi_{\pm
1}|^2$) and spin-density ($\rho_{1} -\rho_{-1}$) profiles of an
individual {\em plus} half-quantum vortex ({\em HQV}) centered at
$r=0$. (b) Interaction potentials (in units of chemical potential
$\mu$) between two {\em HQV}s as a function of separation distance
$d$. $V_{pm}$ is the potential between a {\em plus} {\em HQV} and a
{\em minus} {\em HQV}; $V_{pp}$ is the strong repulsive potential
between two {\em plus} {\em HQV}s. Inset is for spin densities in a
pair of {\em plus}-{\em minus} {\em HQV}s separated at
$d\!\!=\!\!30\xi$, $\xi$ is the healing length.} \label{fig1}
\end{figure}

Our results show that within the core, $|1,1\rangle$ atoms state are
completely depleted while $|1,-1\rangle$ atoms are not. This is
because supercurrents are only present in $|1, 1\rangle$ component;
in fact the density of $|1, -1\rangle$ atoms has an additional small
bump at the center of core, to further take advantage of the
depletion of $|1,1\rangle$ atoms in the same region to minimize the
overall repulsive interactions between $|1,\pm 1\rangle$ atoms. So
the core has a nonzero local spin density $\langle\hat{\bf S}_z({\bf
r})\rangle$($=|\psi_1|^2-|\psi_{-1}|^2$) with excess atoms at
$|1,-1\rangle$ state and we define it as a {\em minus} {\em HQV}.
Similarly, one can construct a {\em plus} {\em HQV} with an
identical vorticity (i.e. $\nabla \times {\bf v}_s$) but with excess
$|1, 1\rangle$ atoms in its core (See Fig.\ref{fig1}(a)). Generally,
{\em HQV}s have distinct spatial magnetic structures: a {\em HQV}
core carries excess spins while far away from the core spin
quadrupole moments slowly rotate around the vortex.

Interactions between two {\em HQV}s very much depend on species of
{\em HQV}s involved. When both are {\em plus} ones (or {\em minus}),
the corresponding interaction potential $V_{pp(mm)}$ is a
logarithmic long range one due to interference between coherent
supercurrents. However, the interaction between a {\em plus} and a
{\em minus} {\em HQV} $V_{pm}$ is repulsive and short ranged only
extending over a scale of vortex cores. In this case, supercurrents
flow in different components and they don't interfere; the short
range interaction is entirely due to inter-component interactions between
$|1,1\rangle$ and $|1,-1\rangle$ atoms. Indeed, the
amplitude of potential $V_{pm}$ is proportional to, when it is small, $c_0-c_2$ which
characterizes mutual interactions between $|1, \pm 1\rangle$ atoms.
An integer vortex or a pair of {\em plus-minus} {\em HQV}s centered
at a same point therefore is unstable and further fractionalizes
into elementary {\em HQV}s. In Fig.\ref{fig1}, we summarize results
of an individual {\em HQV} and two {\em HQV}s.

{\em Dynamical creation of HQVs in rotating BECs} To dynamically
create {\em HQV}s, we numerically solve the time-dependent coupled
GP equations of spin-1 BEC
\begin{eqnarray}
(i-\gamma)\hbar\frac{\partial\psi_{\pm1}}{\partial
t}\!\!&=&\!\![-\frac{\hbar^2}{2m}\nabla^2+V_{tr}-\mu\mp\lambda-\Omega L_z+c_0\rho\nonumber\\
&+&\!\!\!c_2(\rho_{\pm1}+\rho_0-\rho_{\mp1})\!+\!W_\pm]\psi_{\pm1}\!+\!c_2\psi_0^2\bar{\psi}_{\mp1},\nonumber\\
(i-\gamma)\hbar\frac{\partial\psi_0}{\partial
t}\!\!&=&\!\![-\frac{\hbar^2}{2m}\nabla^2+V_{tr}-\mu-\Omega L_z+c_0\rho\nonumber\\
&+&\!\!c_2(\rho_1+\rho_{-1})]\psi_0+2c_2\psi_1\psi_{-1}\bar{\psi}_{0},\label{GPE}
\end{eqnarray}
where $\rho\!=\!\sum_{m_F} \rho_{m_F}$ is the total condensate
density, $V_{tr}({\bf r})$ is a spin-independent confining potential
of an optical trap, and $W_\pm({\bf r})$ are pulsed magnetic
trapping potentials which we further apply in order to create {\em
HQV}s. $\mu$ and $\lambda$ are the Lagrange multipliers used to
preserve the total number and magnetization of atoms respectively;
$\gamma$ is a phenomenological damping parameter which is necessary
for studies of quasi-stationary states \cite{Tsubota02}.

We restrict ourselves to a cigar-shaped potential with the aspect
ratio $\lambda=\omega_\perp/\omega_z\sim14$ which was also used in
early experiments \cite{Madison00}. We consider a two-dimensional
cylindrical trap which is characterized by two dimensionless
parameters $C_0=\frac{8\pi N (a_0+2a_2)}{3L_z}$ and $C_2=\frac{8\pi
N (a_2-a_0)}{3L_z}$, with $L_z$ the size of system along the
$z$-axis and $N=3\times10^6$ the total number of sodium atoms. When
combined with a nonaxisymmetric dipole potential that can be created
using stirring laser beams \cite{Madison00,AboShaeer01}, the optical
trapping potential in a {\em rotating frame} is given by
$V_{tr}({\bf r})=m\omega_\perp^2\{(1+\epsilon) x^2+(1-\epsilon)
y^2\}/2$. Here $\omega_\perp=2\pi\times250$ Hz, and anisotropic
parameter is set to be $\epsilon=0.025$. We also include an
additional magnetic trapping potential: $W_\pm({\bf r})=\mp\beta
m\omega_\perp^2(x^2+y^2)/2$, which could be realized in an
Ioffe-Pritchard trap via a Zeeman splitting $m_Fg_F\mu_B B$ with the
Land\'{e} factor $g_F\!\!=\!\!-1$ for sodium atoms.

We start our numerical simulations with an initial state where $|1,
\pm1\rangle$ are equally populated. Experimentally, it was
demonstrated that when the bias field is small and the gradient
field along the $z$-axis of the trap is almost canceled, such a
state can be prepared and the $|1, \pm1\rangle$ components are
completely miscible (however, with the immiscibility between
$|1,\pm1\rangle$ and $|1,0\rangle$ components) \cite{Stenger98}. We
then study the time evolution of this initial state using the
Crank-Nicolson implicit scheme \cite{Tsubota02}. The unit of length
is $a_h\!\!=\!\!\sqrt{\hbar/2m\omega_\perp}\!\!=\!0.48$ $\mu$m and
the period of the trap is $\omega_\perp^{-1}\!\!=\!4$ ms; the
interaction parameters are $C_0\!\!=\!\!500$ and $C_2\!\!=\!450$ and
the damping rate is $\gamma\!=\!0.03$. Further, we include symmetry
breaking effects in our simulations by allowing the trap center to
randomly jump within a region
$[-\delta,\delta]\!\times\![-\delta,\delta]$ ($\delta\!=\!0.001h$,
where $h$ is the grid size), which is crucial for vortices to enter
the condensate one by one \cite{Parker05} rather than in opposing
pairs \cite{Tsubota02,Lundh03}.

\begin{figure}[t]
\includegraphics[width=\columnwidth]{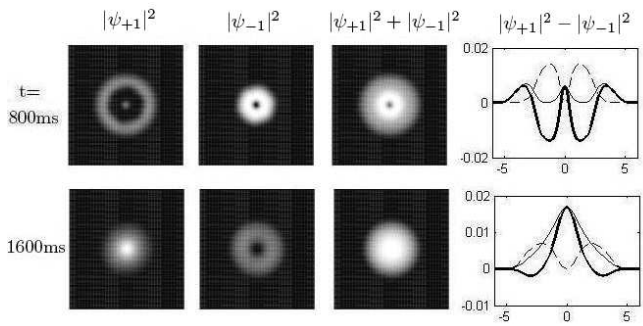}
\caption{ Creation of a half-quantum vortex. Density profiles of
$|\psi_{+1}|^2$, $|\psi_{-1}|^2$,
$|\psi_{+1}|^2\!\!+\!|\psi_{-1}|^2$, and spin-density profile
$|\psi_{+1}|^2$--$|\psi_{-1}|^2$(bold line) are shown. The rotating
frequency is suddenly decreased from an initial value
$\Omega\!\!=\!0.65\omega_\perp$ to $\Omega\!\!=\!0.3\omega_\perp$ at
$t\!\!=\!\!800$ ms. The bottom panel shows that a single
half-quantum vortex is formed at $t\!\!=\!\!1600$ ms after the
magnetic trapping potential has been adiabatically switched off.}
\label{fig2}
\end{figure}

\begin{figure}[t]
\includegraphics[width=\columnwidth]{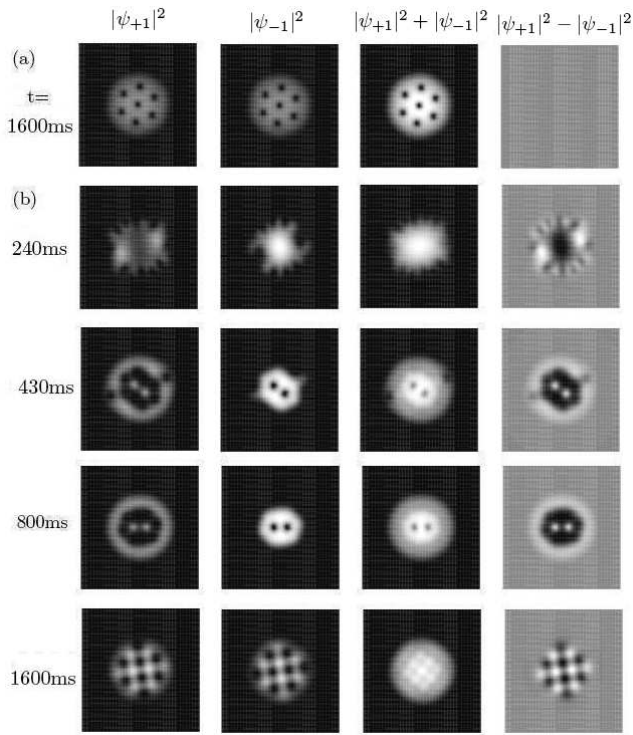}
\caption{ (a) Creation of a triangular integer-vortex lattice in a rotating
optical trap at $t=1600$ ms. (b) Creation of half-quantum
vortex lattices when an additional pulsed magnetic trapping
potential is applied. Here time evolution of various condensate
densities is shown. The optical trap rotates at
$\Omega=0.7\omega_\perp$ with a magnetic trapping potential on until
$t=800$ ms; afterwards, the magnetic trap is adiabatically switched
off within 200 ms. The bottom panel shows the half-quantum vortex
lattice formation at $t=1600$ ms; a square lattice in the
spin-density profile is clearly visible.} \label{fig3}
\end{figure}

However, without additional pulsed magnetic potentials, one can show
that dynamic instabilities for creation of integer vortices in
rotating BECs occur almost at same frequencies as for {\em HQV}s and
a triangular integer-vortex lattice is formed (see
Fig.\ref{fig3}(a)). This integer-vortex lattice is locally stable
with respect to the non-magnetic perturbations, by applying an
additional optical trapping potential with an oscillating trapping
frequency to effectively {\em shake} them, indicating their
metastability. For this reason, a time-dependent magnetic trapping
potential with  harmonic form is applied; and we find that when a
pulsed magnetic field with $\beta > 0.005$ is applied, {\em HQV}
lattices could be formed. Here we set $\beta=0.1$, which is suitable
for both generating a single {\em HQV} and demonstrating the
dynamical evolution of {\em HQV} lattices formation. After the
magnetic trapping potentials are on, $|1,1\rangle$ component further
spreads to the edge, while $|1,-1\rangle$ component remains at the
center of the trap and surfaces of equally populated
$|1,\pm1\rangle$ components become mismatched with two different
Thomas-Fermi radii.

First, we dynamically create a single half-quantum vortex in
condensates, which can be used for the study of dynamics of a {\em
HQV}. We switch on abruptly a rotating drive with
$\Omega=0.65\omega_\perp$ and the trap anisotropy $\epsilon$ is
increased rapidly from zero to its final value 0.025 in 20 ms. At
$t=800$ ms, only one vortex in $|1,-1\rangle$ component appears.
Afterwards, We decrease $\Omega$ to $0.3\omega_\perp$ suddenly, and
switch off the magnetic trapping potential adiabatically within 200
ms. We then find a stable single {\em HQV} formed at  $t=1600$ ms as
shown in Fig.\ref{fig2}. The $0.3\omega_\perp$ frequency used after
$t=1600$ ms is within the stable region estimated earlier
\cite{Isoshima02} and our simulations of dynamics are consistent
with the energetic analysis.

Fractionalized-vortex lattices can be created in a similar setup.
The main experimental procedure and results of our simulations for
creation of {\em HQV} lattices are presented in Fig.\ref{fig3}.
After a rotation with frequency $\Omega=0.7\omega_\perp$ starts
abruptly and anisotropy $\epsilon$ is set to its final value 0.025,
we can see that the cloud is initially elongated and at the same
time rotates with the trap. At about $150$ ms, surface ripples due
to quadrupole excitations occur in $|1,1\rangle$ component of the
condensate, while no surface oscillations appear on the surface of
$|1,-1\rangle$ component. At $t=240$ ms, we find that the density
profile of $|1,1\rangle$ component is along a short-axis while the
$|1,-1\rangle$ component is along a long-axis due to repulsive
interactions between two components, and the surface of
$|1,-1\rangle$ component is not always buried in the inner region of
$|1,1\rangle$. The surfaces of two components oscillate
independently and are decoupled dynamically. At $t=430$ ms, we find
that two {\em plus} {\em HQV}s with excess $|1,1\rangle$ atoms
inside cores have nucleated at the center. Correspondingly, we
observe two small regions near the center where $|1,-1\rangle$ atoms
are completely depleted and the density of $|1,1\rangle$ atoms
remains high. At $t=800$ ms, two components are phase separated, but
the structure of {\em HQV} cores remains almost unchanged. Finally,
we switch off the additional magnetic potential adiabatically within
200 ms and a {\em HQV} lattice with interlaced square configuration
becomes visible. In this structure, to minimize strong repulsive
interactions $V_{pp(mm)}$ between {\em plus} or {\em minus} {\em
HQV}s, the vorticity is evenly distributed among {\em plus} and {\em
minus} {\em HQV}s, or between $|1,\pm 1\rangle$ components. This
also indicates that spatially each {\em plus} {\em HQV} prefers to
be adjacent to {\em minus} {\em HQV}s and {\em vice versa} to avoid
stronger interactions $V_{pp}, V_{mm}$ and to take advantage of
relative weaker interactions $V_{pm}$ (See Fig.\ref{fig1}). To
further minimize repulsive interactions $V_{pm}$ between nearest
neighboring vortices, a {\em plus} {\em HQV} is displaced away from
adjacent {\em minus} {\em HQV}s by a maximal distance. Generally
because of the asymmetry between $V_{pp}$ and $V_{pm}$, a bipartite
vortex lattice should be favored over frustrated geometries such as
triangular lattices where a {\em plus} {\em HQV} could be adjacent
to another {\em plus} {\em HQV} resulting in much stronger
repulsion. In our simulations of $^{23}$Na atoms in rotating traps
with $c_0 \simeq30 c_2$, square vortex lattices have always been
found. Equilibrium energetics of rectangular or square lattices were
also considered in the quantum Hall regime \cite{Kita02,Mueller02},
in two-component BECs coupled by an external driving field where
vortex molecules are formed \cite{Kasamatsu03}, and also observed in
condensates of pseudo-spin-$1/2$ rubidium atoms
\cite{Schweikhard04}. Here we have mainly focused on {\em dynamical
creation} of {\em HQV} lattices confined to a spin-density-wave
structure at relatively low frequencies; this structure can be
conveniently probed by taking absorption images of ballistically
expanding cold atoms in a Stern-Gerlach field \cite{Stenger98}.

In conclusion, we have demonstrated a practical setup to create
fractionalized vortices and vortex lattices in BECs of sodium atoms.
We found that a {\em square} half-quantum vortex lattice has a
distinct periodically modulated spin-density-wave spatial structure,
due to short range repulsive interactions between neighboring
half-quantum vortices. Our results are of particular significance
for creating these excitations in experiments and for exploring
novel phenomena associated with them.

FZ would like to thank E. Demler, W. Ketterle, K. Madison for
stimulating discussions. This work is supported by the office of the
Dean of Science, UBC, NSERC (Canada), Canadian Institute for
Advanced Research, A. P. Sloan foundation; NSFC under grant
90406017, 60525417, the NKBRSFC under grant 2006CB921400.

\end{document}